\documentclass[aip,jcp,amsmath,amssymb,reprint]{revtex4-1}
\usepackage{braket,nicefrac}
\usepackage{graphicx}
\usepackage{times}
\usepackage{hyperref}

\newcommand{\eqspace}[1]{\text{\hspace{#1\textwidth}}}

\begin{document}
\title{A two-density approach to the general many-body problem and a proof of principle for small atoms and molecules}
\author{Thomas Pope}
\email{Thomas.Pope2@newcastle.ac.uk}
\affiliation{School of Natural and Environmental Sciences, Newcastle University, Newcastle NE1 7RU, United Kingdom}
\author{Werner Hofer}
% \email{Werner.Hofer@newcastle.ac.uk}
\affiliation{School of Natural and Environmental Sciences, Newcastle University, Newcastle NE1 7RU, United Kingdom}
\affiliation{Institute of Physics \& University of Chinese Academy of Sciences, Chinese Academy of Sciences, Beijing 100190, China.}
% \date{\today}
\begin{abstract}
 An extended electron model fully recovers many of the experimental results of quantum mechanics while it avoids many of the pitfalls and remains generally free of paradoxes. The formulation of the many-body electronic problem here resembles the Kohn-Sham formulation of standard density functional theory. However, rather than referring electronic properties to a large set of single electron orbitals, the extended electron model uses only mass density and field components, leading to a substantial increase in computational efficiency. To date, the Hohenberg-Kohn theorems have not been proved for a model of this type, nor has a universal energy functional been presented. In this paper, we address these problems and show that the Hohenberg-Kohn theorems do also hold for a density model of this type. We then present a proof-of-concept practical implementation of this method and show that it reproduces the accuracy of more widely used methods on a test-set of small atomic systems, thus paving the way for the development of fast, efficient and accurate codes on this basis.
\end{abstract}
\maketitle

\section*{Introduction}
Orbital-free density functional theory has the capacity to vastly increase the computational efficiency of electronic structure calculations~\cite{levy1979universal,*levy1984exact,pearson1993ab,wesolowski2013recent,lehtomaki2014orbital,karasiev2015frank}. However, a general kinetic energy functional has remained elusive~\cite{garcia2008approach,huang2010nonlocal,shin2014enhanced,mi2018nonlocal,constantin2018nonlocal,*constantin2018semilocal}. In early formulations, the kinetic energy was approximated by an interpolation between the Thomas-Fermi functional and the von Weizs\"{a}cker-type gradient expansion. In modern density functional theory (DFT) methods this typically requires the implementation of the Kohn-Sham (KS) method~\cite{kohn1965self}, which involves decomposing the density into a set of single electron orbitals. As a result, electronic structure simulations are significantly more expensive computationally. Recent improvements in the efficiency have been made using massively parallel approaches and highly optimized algorithms~\cite{jones2015density}. Sparse matrix techniques~\cite{michaud2016rescu} have been employed to reduce the scaling significantly and various order-N approaches~\cite{soler2002siesta,skylaris2005introducing} have introduced linear scaling DFT to systems with non-zero band gaps. In all these methods, however, the fundamental need to treat each electron separately has not been overcome. 

Recently, orbital-free DFT has been shown to be tractable with machine-learning routines by the Burke group~\cite{brockherde2017bypassing}. However, in this case, the limitation is that the method so far only works for a distinct number of atoms and a distinct number of electrons. Removing the restrictions on these numbers, making such an approach truly general, still seems to be beyond current methods.

An extended electron model, by contrast, reformulates the many-body problem for a general fermionic distribution of charge and spin on the basis of real and extended electrons, so the ensuing mathematical formalism is linearly scaling from the outset~\cite{hofer2011unconventional}. The method fully recovers the formulation of quantum mechanics in terms of the Schr\"{o}dinger equation~\cite{hofer2011unconventional} and avoids many of the paradoxes that arise from the Copenhagen interpretation~\cite{pope2017spin}, but so far it has not been shown that the Hohenberg-Kohn (HK) Theorems~\cite{hohenberg1964inhomogeneous}, which underpin DFT, apply also to a system composed of unique mass and spin densities. Here, we present a short introduction to the extended electron framework in the context of many-body electronic structure calculations and arrive at a set of equations that are formally equal to the KS equations of DFT. But while the KS approach entails a sum over many single electron orbitals, the extended electron approach is simplified to a sum over two densities. Subsequently, we present a general formulation of the many-body problem and show that the HK theorems are satisfied. Finally, a proof-of-concept implementation is presented and shown to reproduce the results of other methods for a test-set of small atomic systems. 

\section*{An Extended Electron Model}
There are many attempts to model an electron as something other than a point particle with an intrinsic momentum in the literature. For example the approaches by de Broglie-Bohm~\cite{de1927wave,*bohm1952suggested,*bohm1952suggestedII} or Hestenes~\cite{hestenes1973local,*hestenes1985quantum,*hestenes1990zitterbewegung}, where the electron has been linked to a field-like construct. But applying a strict physical reality to the wavefunction, while retaining the mechanical properties of a point particle, seems to lead inevitably to non-local potentials~\cite{bell1966problem}. Moreover, recent advances in the precision in STM measurements~\cite{rieder2004scanning} have been shown to violate the Heisenberg uncertainty relations~\cite{hofer2012heisenberg}, which makes it difficult to retain the view that the density is a statistical quantity. But if it is not a statistical quantity, then the electron must be an extended object in space. In the proposed extended electron model we relax the assumption that the electron is a point particle and instead model it as an extended density distribution from the very outset, much in line with the common practice in DFT, but with the extension into the domain of single electrons. In this context, the densities measured in STM experiments are not probability densities, but physically real number densities. The wave properties observed, rather than being properties of an underlying wavefunction, are then simply real physical properties of electrons in motion and are contained in spatially extended density oscillations. 
 
\begin{figure}[!ht]
  \centerline{\includegraphics[width=.85\linewidth]{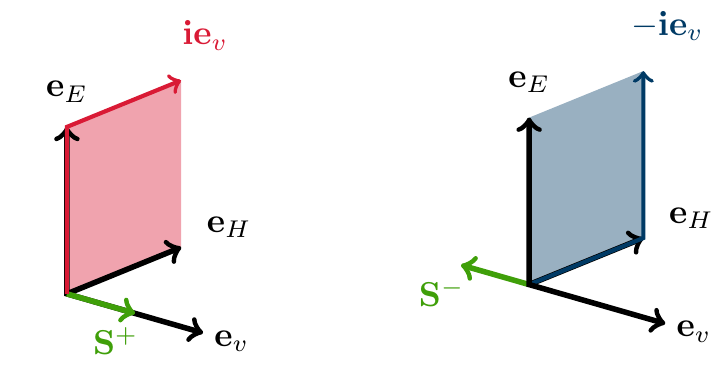}}
  \caption{Schematic of electron spin and field vectors (${\bf{S}}$, ${\bf{e}}_{E}$, and ${\bf{e}}_{H}$,respectively) and the vector of motion, ${\bf{e}}_{v}$, for an electron, with both the parallel (left) and antiparallel (right) behaviour. The direction of the electron spin vector, ${\bf{S}}^{\pm}$, is shown by the short arrows on the ${\bf{e}}_{v}$ axis in both cases.}
  \label{field_vectors}
 \end{figure}

 Energetically, these oscillations are supplemented by equal and opposite oscillations in a second density parameter associated with the field components. In order to model this behaviour, we use the framework of geometric algebra~\cite{hestenes2012clifford,gull1993imaginary}. We postulate that an electron has intrinsic electromagnetic potentials associated with the vector of motion, ${\bf{e}}_{v}$. The direction of the electric, $\text{\bf{E}}$, and magnetic, $\text{\bf{H}}$, fields are given by unit vectors, ${\bf{e}}_{E}$ and ${\bf{e}}_{H}$ respectively, that are perpendicular to one another and to the vector of motion (see figure~\ref{field_vectors}). Within the framework of geometric algebra, the product of these two vectors is a bivector, with anticommutative properties, ${\bf{e}}_{H}{\bf{e}}_{E}=-{\bf{e}}_{E}{\bf{e}}_{H}$. Finally, the product of three perpendicular vectors is the so-called pseudoscalar, ${\bf{i}}$, which when multiplied by a vector gives a bivector plane perpendicular to that vector, ${\bf{i}}{\bf{e}}_{v}={\bf{e}}_{E}{\bf{e}}_{H}$. We may now define an effective spin vector, ${\bf{e}}_{S}$, which multiplies with the pseudoscalar to give the bivector plane of the two electromagnetic potentials with either the clockwise or anti-clockwise \emph{handedness}. Thus the spin vector may be parallel or anti-parallel to the vector of motion.  An additional phase must also be added to account for the energy conservation of electrons at the local level~\cite{hofer2011unconventional}, but for simplicity, we set this term to zero in our notation.
 
 We define a Poynting-like vector, consisting of the the spin density and the spin unit vector, ${\bf{i}}
  {\bf{e}}_{S}S=\text{\bf{E}}\text{\bf{H}}$. The spin density is the energy density of the field components of the electron. Finally, the effective wavefunction is a multivector~\cite{doran2003geometric} given in terms of the mass density, $\rho$, the spin density and the direction of the spin vector,
 \begin{equation}
  \Psi\!\left({\bf{r}}\right)=\rho^{\nicefrac{1}{2}}\!\left({\bf{r}}\right)+{\bf{i}}
  {\bf{e}}_{S}\!\left({\bf{r}}\right) S^{\nicefrac{1}{2}}\!\left({\bf{r}}\right).
 \end{equation}

 The duality operation, $\Psi^{\dagger}$, is represented by a change in the helicity of the bivector term and, hence, a change in sign of the spin vector. The product of $\Psi$ and $\Psi^{\dagger}$ complies with the Born rule and corresponds to the inertial number density of the electron, $\Psi\Psi^{\dagger}=\rho+S=\rho_{0}$. Thus, the wave properties are related to oscillations in the mass density of the electron, which are supplemented by equal and opposite oscillations in the spin density, $\dot{S}=-\dot{\rho}$.  
 
 A model of this type has several advantages compared to the standard model of a point-like electron, even before the marked improvement in computational efficiency is considered. Crucially, it is built from measurable components - namely mass density and field components. This provides more insight and leads to a higher level of scrutiny than the fundamentally immeasurable wavefunction of the Copenhagen interpretation. For example, while in the standard approach, spin is modelled using a spinor operated on by the Pauli matrices~\cite{benenti2004principles}, in the extended electron model, the spin vector is contained in the bivector term of the wavefunction and, within the framework of geometric algebra, the bivector components automatically reproduce Pauli algebra~\cite{doran1993states}. While the model explicitly contains the spin as a vector variable, it is configured mathematically in such a way as to enable us to maintain isotropy with respect to rotations in the bivector plane perpendicular to that vector. Since this direction is due to the motion of the electron, a statistical manifold of an equal number of spin-up and spin-down electrons will remain fully isotropic. This allows to reproduce the results of Stern-Gerlach type experiments~\cite{hofer2011unconventional,hofer2014elements} and even resolve the Einstein-Podolsky-Rosen paradox~\cite{pope2017spin,hofer2012solving}.
 
 \section*{Many-Body Problem}
 In a general system composed of many electrons the Hamiltonian, within the extended electron model, is composed of four terms, which respectively represent the kinetic energy, the fixed external field due to the nuclei, the combined Hartree and cohesive potential and the bivector potential~\cite{hofer2011unconventional},
 \begin{equation}
  H=-\frac{1}{2}\nabla^{2}+v_{\text{ext}}+v_{\text{eff}}+
  {\bf{i}}{\bf{e}}_{b}v_{b}=\hat{h}_{0}+{\bf{i}}{\bf{e}}_{b}v_{b}.
 \end{equation}

 The kinetic term, the Hartree-cohesive potential and the bivector potential are all universal potentials given by the density properties of the electron alone. The Hartree-cohesive potential is due to Coulomb interaction. The bivector potential concerns the field-mediated contribution, which includes any electron-electron interaction not captured by the Hartree-cohesive potential - for example, the exchange potential. And the kinetic energy is given by the Laplace operator acting on the mass and spin densities. The only system-specific potential is the external potential due to the nuclei, which is defined in the same way as in standard DFT methods. For simplicity, we collect all but the bivector potential into a Hamiltonian operator, $\hat{h}_{0}$ so that the Schr\"{o}dinger equation is given by,
 \begin{equation}
  \left(\hat{h}_{0}+{\bf{i}}{\bf{e}}_{b}v_{b}\right)\Psi=
  \mu\Psi\label{eese}
 \end{equation}
  
 We may decompose the equation into scalar and bivector components, considering that the Laplacian is a scaler differential operator and that the geometric product of vectors yields a scalar dot product and a bivector wedge product,
 \begin{align}
  \hat{h}_{0}{\bf{i}}{\bf{e}}_{S}S^{\nicefrac{1}{2}}-
  {\bf{e}}_{b}\wedge{\bf{e}}_{S}v_{b}S^{\nicefrac{1}{2}}+
  {\bf{i}}{\bf{e}}_{b}v_{b}\rho^{\nicefrac{1}{2}}&=
  \mu{\bf{i}}{\bf{e}}_{S}S^{\nicefrac{1}{2}},\\
  \hat{h}_{0}\rho^{\nicefrac{1}{2}}-{\bf{e}}_{b}\cdot{\bf{e}}_{S}
  v_{b}S^{\nicefrac{1}{2}}&=\mu\rho^{\nicefrac{1}{2}}.
  \label{coupledse}
 \end{align}
 
 The difference between this formalism and that of standard DFT is not simply a potential term of the same general form as the effective or external potential. In the scalar part, we have an additional term which will act as a source for the electron density distribution. This source term depends on the direction of the bivector potential as well as the field components of electron motion. Thus, the bivector part introduces directional effects into the density equation, and couples density and field components. Strictly speaking, we expect that a Galilean boost factor will also be needed to deal with the problem of Galilean invariance - similar to the boost factor used in the Schr\"{o}dinger equation~\cite{hamermesh1960galilean}. For simplicity, we assume a static frame and set this additional term to zero.
 
 We note that the two coupled equations only have a solution if the direction of the spin is parallel to the direction of the bivector potential, ${\bf{e}}_{S}={\bf{e}}_{b}$, or if the direction of the bivector is antiparallel, in which case the sign of the potential is reversed. This feature is also present in standard formulations of the problem; the interaction with external fields breaks the isotropy of spin. However, while exchange interactions are not directional from the outset, the bivector potential clearly is, which indicates that there is a more direct connection in the present framework between magnetic properties and magnetic interactions. Here, the existence of a directional crystal field forces alignment of the electron spin. 
 
 Finally, we arrive at a set of equations that are formally equal to the KS equations in DFT, but while the KS equations describe single electron orbitals, these directly describe the square root of the mass and spin densities, 
 \begin{equation}
  \begin{pmatrix}\hat{h}_{0}&-v_{b}\\v_{b}&\hat{h}_{0}
  \end{pmatrix}\begin{pmatrix}\rho^{\nicefrac{1}{2}}\\
  S^{\nicefrac{1}{2}}\end{pmatrix}=\mu\begin{pmatrix}
  \rho^{\nicefrac{1}{2}}\\S^{\nicefrac{1}{2}}\end{pmatrix}.
  \label{kohnshameq0}
 \end{equation}
 The bivector potential serves to tie the two density equations together. If we set the potential to zero, the two densities are essentially independent. 
 
 Multicomponent generalizations of DFT have existed almost as long as DFT itself.  The most widely used is, of course, spin-polarized DFT, where electrons of different spin are modelled using separate density terms. But it has also been used to model electron-hole drops in semiconductors~\cite{sander1973surface,kalia1978surface}, where both quantum objects are modelled using a set of KS equations that interact via a potential. Finally, it can be used to model the electron-positron interaction~\cite{boronski1986electron} by, again, treating the two physical objects with two sets of KS equations. In all cases, the need to treat the densities as a sum of single-object wavefunctions is still present. Multicomponent DFT has also been used to relax the Born-Oppenheimer approximation by modelling both the electrons and the nuclei as quantum objects with associated density parameters~\cite{kreibich2001multicomponent,*kreibich2008multicomponent}. Again, the electrons are decomposed into single-electron wavefunctions for accuracy. In all cases, these examples seek to model two distinct quantum objects using two density parameters, while in the extended electron model, two physical properties of a single quantum object are being modelled. In addition, the examples typically adhere to the KS implementation of DFT, with the obvious caveat that there are now two interacting sets of KS equations. The need to treat the densities as a sum of single-object wavefunctions is still present. By contrast, the extended electron model replaces the KS equations entirely with a simple two-density problem.

 Here, we introduce a form for the kinetic energy that is similar to the von Weizs\"{a}cker functional, where rather than acting on the sum of the densities, the Laplace operator acts on each density individually,
 \begin{equation}
  T_{\mathcal{X}}=\mathcal{X}^{\nicefrac{1}{2}}
  \left(-\frac{1}{2}\nabla^{2}\right)\mathcal{X}^{\nicefrac{1}{2}}
 \end{equation}
 We then multiply the two density equations by the square root of either the mass or spin density respectively,
 \begin{alignat}{4}
  &T_{\rho}+&\left(v_{\text{ext}}+v_{\text{eff}}-\mu\right)
  \rho-&v_{b}\left(S\rho\right)^{\nicefrac{1}{2}}&=0,\\
  &T_{S}+&\left(v_{\text{ext}}+v_{\text{eff}}-\mu\right)S+
  &v_{b}\left(S\rho\right)^{\nicefrac{1}{2}}&=0,
 \end{alignat}
 By summing the two equations, or subtracting the latter from the former, we find respectively
  \begin{align}
   T_{\rho}+T_{S}+\left(v_{\text{ext}}+v_{\text{eff}}-
   \mu\right)\left(\rho+S\right)&=0,\label{kohnshameq1}\\
   T_{\rho}-T_{S}+\left(v_{\text{ext}}+v_{\text{eff}}-
   \mu\right)\left(\rho-S\right)&=2v_{b}
   \left(S\rho\right)^{\nicefrac{1}{2}},\label{kohnshameq2}
  \end{align}

 The first equation yields an expression for the chemical potential, which when substituted into the second gives an expression for the bivector potential that depends solely on the mass and spin densities, 
 \begin{equation}
  v_{b}\left(\rho+S\right)=S^{\nicefrac{1}{2}}\left(
  -\frac{1}{2}\nabla^{2}\right)\rho^{\nicefrac{1}{2}}
  -\rho^{\nicefrac{1}{2}}\left(-\frac{1}{2}\nabla^{2}
  \right)S^{\nicefrac{1}{2}},\label{biveq}
 \end{equation}
 and substituting this into Eq.~\ref{kohnshameq2}, we find,
 \begin{equation}
  \frac{\rho-S}{\rho+S}\left(T_{\rho}+T_{S}\right)+
  \left(v_{\text{ext}}+v_{\text{eff}}-\mu\right)\left(\rho-
  S\right)=0\label{kohnshameq3}
 \end{equation}

 The main components of the bivector potential are the crossover terms between the mass and spin densities and their differentials. Note that the potential is only non-zero when the mass and spin densities are not equal, which agrees with expectation since the wave properties of the electron are due to equal and opposite fluctuations in these densities. In this sense, the bivector potential is a representation of this quantum behaviour while the Coulombic components are contained in the effective and external potentials. 
 
 The effective potential contains both the Hartree term and a cohesive term to account for self interaction. In principle, the cohesive term makes sure that this potential vanishes in the single electron case. However, in larger systems an approximation is needed. Removing the self-interaction energy will in all likelihood be one of the more complicated problems in the implementation of the extended electron model. Approaches often involve an orbital-by-orbital subtraction of the self-interaction~\cite{perdew1981self} or by including terms like the Fock exchange so that the self interaction terms cancel with their equivalents in the Hartree potential~\cite{paier2005perdew}. Both methods are prohibitively expensive. As a first step, we employ the local density approximation (LDA). The effective potential is the sum of the Hartree and cohesive potentials, $v_{\text{eff}}=v_{\text{H}}+v_{\text{coh}}$, where the Hartree potential is described conventionally,
\begin{equation}
 v_{\text{H}}\!\left({\bf{r}}\right)=\int{\frac{\text{n}\left(
 {\bf{r}}'\right)}{\left|{\bf{r}}-{\bf{r}}'\right|}\text{d}^{3}
 {\bf{r}}'},
\end{equation}
and the cohesive potential is described by assuming the electron density is slowly varying. In this case, a single electron occupies a sphere with a Wigner-Seitz radius of $r_{s}$, such that,
\begin{equation}
 \frac{4\pi}{3}r_{s}^{3}\text{n}\left({\bf{r}}\right)=1\implies 
 r_{s}=\left(\frac{4\pi}{3}\text{n}\left({\bf{r}}\right)\right)
 ^{-\nicefrac{1}{3}}.
\end{equation}
In the approximation, the cohesive potential is equal and opposite to the repulsive Hartree contribution,
\begin{equation}
 v_{\text{coh}}^{\text{LDA}}=-\text{n}\left({\bf{r}}\right)
 \int_{0}^{r_{s}}{\frac{4\pi}{3}r^{3}\cdot\frac{1}{r}\cdot4\pi 
 r^{2}\text{d}r}=-\frac{3}{5}r_{s}^{-1}.
\end{equation}

The external potential will contain the effect of the nuclei and, for computational efficiency, the effect of the core electrons. This is achieved by constructing pseudopotentials~\cite{hamann1979norm}, which are generated from all-electron atomic calculations~\cite{perdew1986accurate,pople1992kohn,white1994implementation} by assuming spherical screening and self consistently solving the radial KS equation~\cite{hohenberg1964inhomogeneous,kohn1965self}. From the single atom solutions, we calculate the pseudopotential that replaces the effect of the core electrons.

\section*{Hohenberg-Kohn Theorems with Extended Electrons}
 The HK theorems underpin modern DFT. In the original formulation, the two theorems assert that, firstly, the particle density is provably unique for any unique external potential and, secondly, the ground state density minimises the energy. Since the formulation for extended electrons closely resembles the KS equations, it is clear that the second theorem holds. Here, we concentrate on the first, which is essentially a proof that there is a unique map between the ground state density and the potential,
 \begin{equation}
  v_{\text{ext}}\Longleftrightarrow\rho+S.
 \end{equation}
 
 It was shown soon after the HK theorems were published that they do not apply to spin-polarized DFT~\cite{von1972local} and, instead, the proof entails two unique maps; one between the ground state density and the wavefunction and the other between the wavefunction and the 
 potential,
  \begin{equation}
  v_{\text{ext}}\Longleftrightarrow\Psi\Longleftrightarrow\rho,S.  
 \end{equation}
 
 The second of these maps, between the potential and the wavefunction, is essentially guaranteed by the Schr\"{o}dinger equation. The first map, between wavefunction and density, is necessarily guaranteed in the extended electron model since the wavefunction is constructed from the densities by definition. Here we show that, within the framework of extended electrons, it is possible to circumvent the wavefunction and create a direct map between the densities and the potential. We begin by considering Eq.~\ref{kohnshameq1}, which when integrated yields the total energy of the system. By definition, the density that minimises the total energy is the ground state and so any other density will yield a higher total energy. We fix the mass and spin densities and calculate the total energy for two potentials, $v_{\text{ext}}$ and $v_{\text{ext}}'$. The difference between the two energies is positive by definition, and, because all terms excluding the external potential are defined from the densities, they cancel to give,
 \begin{equation}
  \int{\left(v_{\text{ext}}'-v_{\text{ext}}\right)\left(\rho+S\right)
  \text{d}^{3}{\bf{r}}}=\delta\mu\int{\left(\rho+S\right)
  \text{d}^{3}{\bf{r}}}>0.
 \end{equation}

 We now assume that the mass density is unchanged, but the spin density is altered so that it minimises the second potential rather than the first,
 \begin{equation}
  \int{\left(v_{\text{ext}}-v_{\text{ext}}'\right)
  \left(\rho+S'\right)\text{d}^{3}{\bf{r}}}=\delta\mu'
  \int{\left(\rho+S'\right)\text{d}^{3}{\bf{r}}}>0.
 \end{equation}
 Summing the two leads to the first condition we must satisfy,
 \begin{align}
  &\int{\left(v_{\text{ext}}'-v_{\text{ext}}\right)
  \left(S-S'\right)\text{d}^{3}{\bf{r}}}
  \nonumber\\&\eqspace{0.08}=\delta\mu
  \int{\left(\rho+S\right)\text{d}^{3}{\bf{r}}}+\delta
  \mu'\int{\left(\rho+S'\right)\text{d}^{3}{\bf{r}}}\label{con1}
 \end{align}
 
 Next we consider Eq.~\ref{kohnshameq3}, which, if we minimize the first potential, gives,
 \begin{equation}
  \int{\left(v_{\text{ext}}'-v_{\text{ext}}\right)
  \left(\rho-S\right)\text{d}^{3}{\bf{r}}}=\delta\mu
  \int{\left(\rho-S\right)\text{d}^{3}{\bf{r}}}.
 \end{equation}
 and if we adjust the spin density to minimise the second potential, gives, 
 \begin{equation}
  \int{\left(v_{\text{ext}}-v_{\text{ext}}'\right)
  \left(\rho-S'\right)\text{d}^{3}{\bf{r}}}=\delta
  \mu'\int{\left(\rho-S'\right)\text{d}^{3}{\bf{r}}}.
 \end{equation}
 
 The sum of these two equations gives us a second condition,
 \begin{align}
  &\int{\left(v_{\text{ext}}'-v_{\text{ext}}\right)
  \left(S'-S\right)\text{d}^{3}{\bf{r}}}
  \nonumber\\&\eqspace{0.08}=\delta\mu
  \int{\left(\rho-S\right)\text{d}^{3}{\bf{r}}}+
  \delta\mu'\int{\left(\rho-S'\right)\text{d}^{3}{\bf{r}}}\label{con2}
 \end{align}

 Finally, by summing the two conditions we find that the only case for which the mass density can remain constant for two unique external potentials is described by,
 \begin{equation}
  \left(\delta\mu+\delta\mu'\right)
  \int{\rho\text{\hspace{0.005\textwidth}}
  \text{d}^{3}{\bf{r}}}=0.
 \end{equation}
 
 Similarly, by fixing the spin density and following the same procedure\footnote{In this case, rather than summing equations~\ref{con1} and \ref{con2}, we subtract one from the other.} we find,
 \begin{equation}
  \left(\delta\mu+\delta\mu'\right)
  \int{S\text{\hspace{0.005\textwidth}}
  \text{d}^{3}{\bf{r}}}=0.
 \end{equation}
 
 In other words, the only way the same mass or spin density can minimise two external potentials is if it is zero throughout space, in which case the two-density problem is reduced to a single density equation, which necessarily obeys the HK theorems. This has been noted in the case of spin-polarized DFT, where when the spin is fully polarized, one density is zero and hence the associated potential is undefined~\cite{eschrig2001density,gidopoulos2007potential}. In this case, the second potential is unimportant since it only acts of the zeroed density and, since a central argument of the extended electron model is that the wave properties of the electron are manifest as oscillations in the mass density, which are then energetically supplemented by the spin density, the scenario in which either density is zero throughout space is not physically relevant. 
  
\section*{Implementation}
 In order to test this model on a range of systems, we have partially implemented it into a development version of the CASTEP implementation of DFT~\cite{clark2005first} as an alternative density-mixing algorithm. Similar to the auxiliary functional approach~\cite{hasnip2015auxiliary}, the model takes the KS density at each stage of the self-consistency cycle, independently minimises it according to the extended electron many-body equations and returns a density to the parent KS cycle. Convergence occurs when consecutive densities returned by the extended electron model are equivalent and thus the results of the two models are consistent. Since correlation is not yet present in the extended electron model, we employ an LDA correlation term~\cite{perdew1981self}. 
 
 In principle, this approach should lead to an improvement in performance due to the reduction in the number of cycles through the KS section of the algorithm. The extended electron approach is expected to provide a density much closer to the ground state than traditional mixing routines. However, an exact mapping between the spin-polarized KS DFT and the extended electron model does not exist and is beyond the scope of this paper. Instead, we present the approach as a proof-of-concept implementation.
 
 While comparing the ground state densities of two approaches has been used effectively~\cite{medvedev2017density}, it is entirely dependent on the choice of metric~\cite{burke2018quantifying}. So, rather than comparing the densities directly, we present all physical properties as calculated by existing CASTEP routines. The density, minimised under the KS protocol or by the extended electron model, is used to calculate the total energy using the KS method with the LDA functional. In all calculations, a cut-off energy of 10 keV is used. We also employ \emph{on-the-fly} pseudopotentials calculated using the KS formalism to incorporate the core electrons~\cite{vanderbilt1990soft}. For a direct comparison, we use the difference in energy, $\Delta\text{E}$, calculated by subtracting the total energy for the extended electron density from the total energy for the KS density.
 
\begin{figure}%[!ht]
 \centerline{\includegraphics[width=\linewidth]{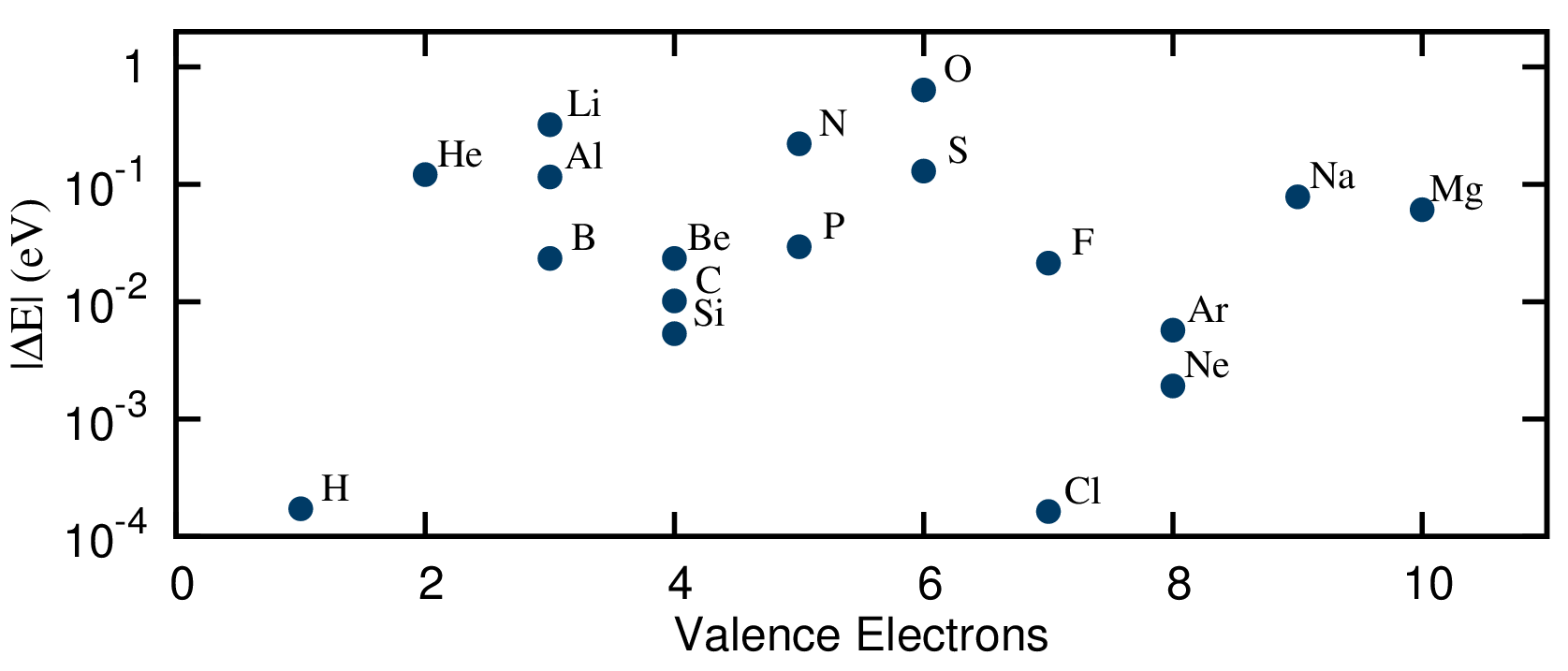}}
 \caption{Difference in energy between the extended electron model and the KS approach for a range of single atom systems against the number of valence electrons in the calculation.}
 \label{atoms}
 \end{figure}

 In figure~\ref{atoms}, we show the energy differences calculated in single-atom calculations for the first two rows of the periodic table. We find good agreement between the models and, importantly, no systematic error emerging as the number of valence electrons is increased. While we find the discrepancy is always less than 0.05\% of the total energy, with an average of $\left(-2.220\pm0.003\right)\times10^{-5}\times\text{E}_{\text{TOT}}^{\text{KS}}$, we expect further improvement with the future implementation of bespoke pseudopotentials developed self-consistently from the extended electron formalism. In addition to increased accuracy, this is also expected to improve efficiency. 

\begin{figure}%[!ht]
 \centerline{\includegraphics[width=\linewidth]{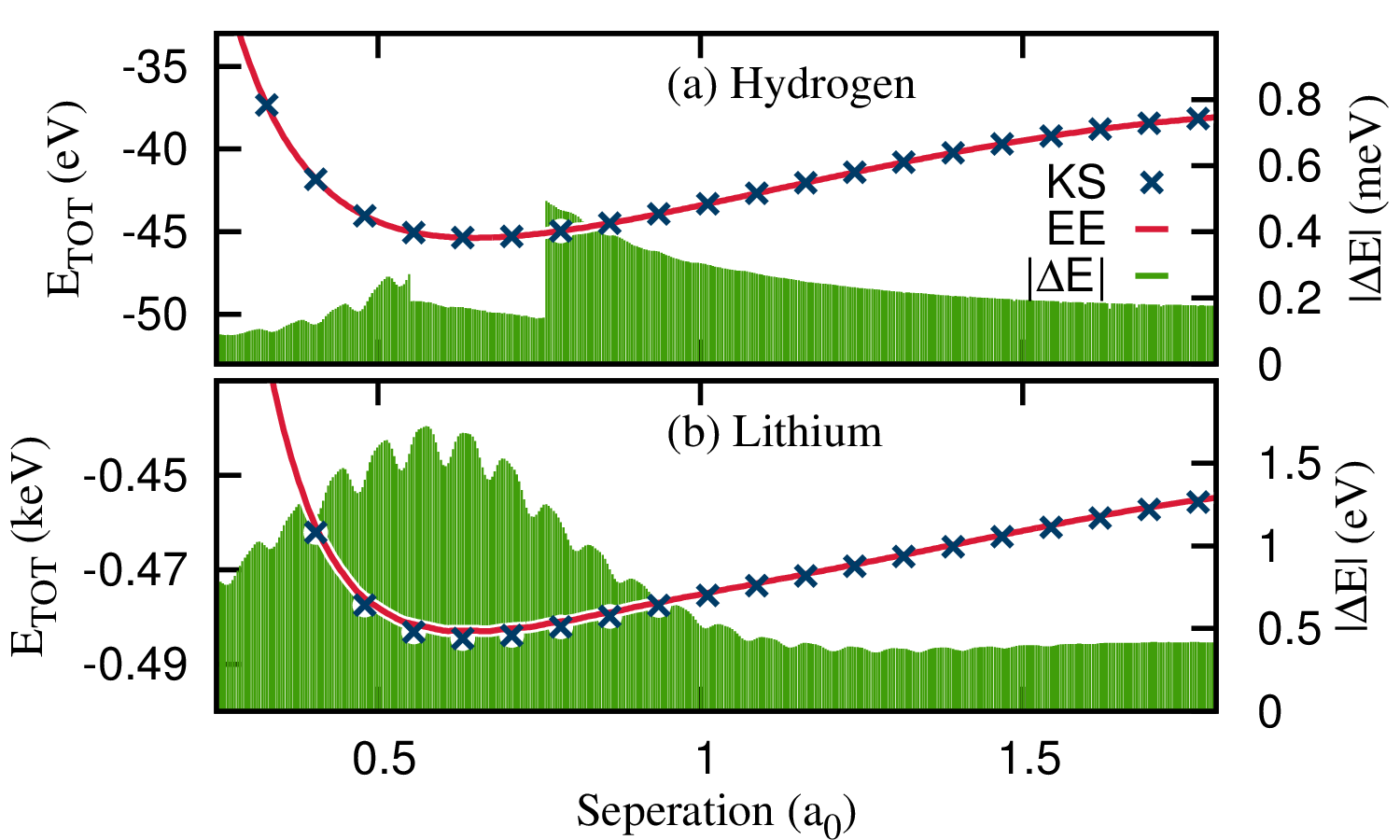}}
 \caption{Energy as a function of bond length, calculated using the extended electron and Kohn-Sham methods, for the Hydrogen (a) and Lithium (b) dimers. The absolute energy difference between the two models is plotted for each calculation.}
 \label{dimers}
 \end{figure}

 In figure~\ref{dimers} we show the calculated total energy as a function of bond length for two sample dimer molecules (Hydrogen and Lithium) and find good agreement between the two models. In particular, while the energies vary by around 0.1\% of the total energy, the predicted bond lengths remain almost identical. 

\begin{figure}%[!ht]
 \centerline{\includegraphics[width=\linewidth]{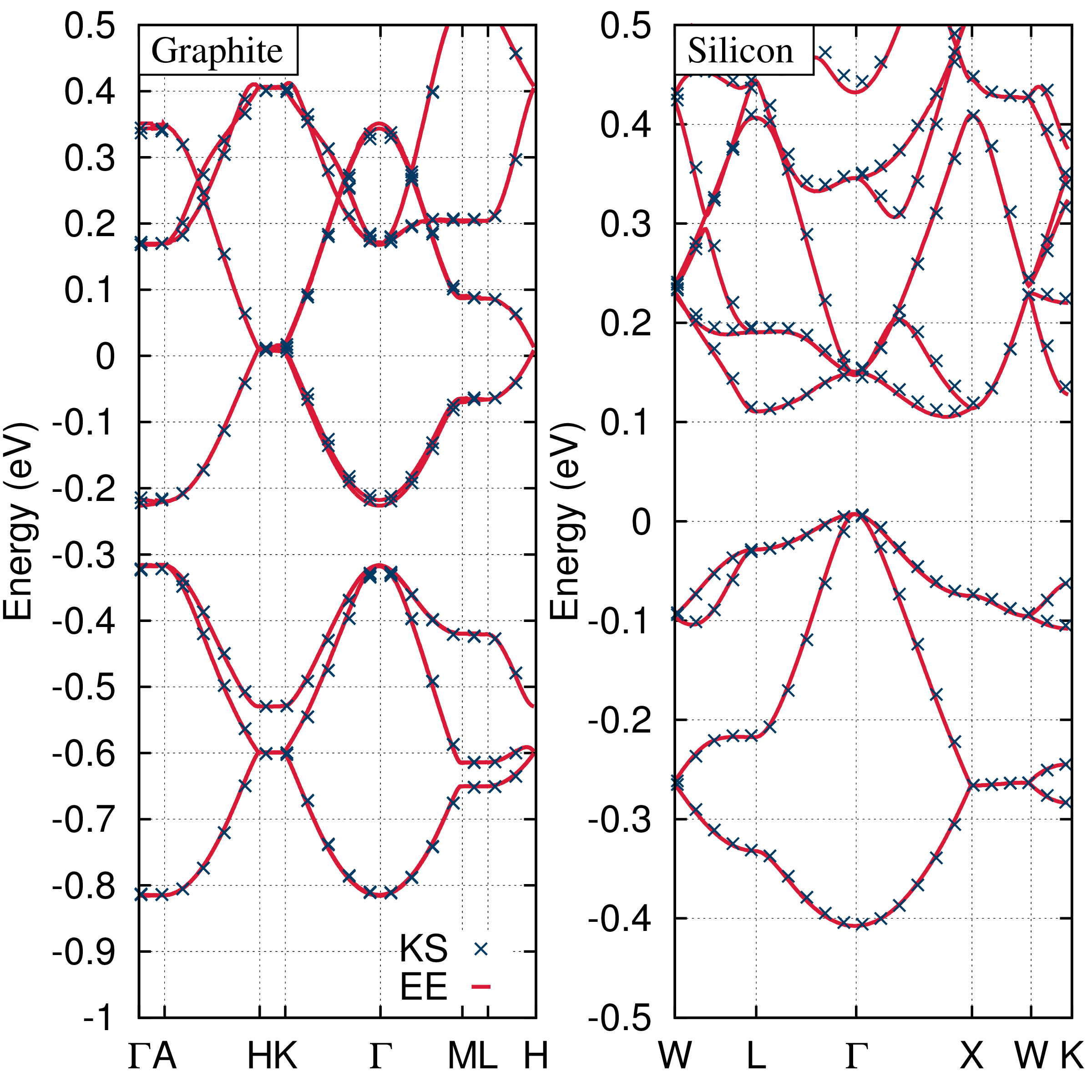}}
 \caption{Bandstructure for graphite (left) and silicon (right) as calculated using the extended electron method (red) and KS method (blue).}
 \label{bands}
 \end{figure}

 Finally, we calculate the ground state density for two crystal structures (graphite and silicon) using both the extended electron and the KS methods and use CASTEP's inbuilt routines to evaluate the band structure. We plot the results in figure~\ref{bands} and see very good agreement between the two methods. Typically, the largest discrepancies happen in the highest energy levels, with the occupied states all conforming to expectation. In addition, both methods predict the same band-gaps for both crystals. 

\section*{Conclusion}
 In conclusion, we have presented an extended electron approach to the many-body problem and have shown that the HK theorems remain valid for the two-density formulation. We have formulated a many-body approach for describing the ground-state energy of the system, which encodes the exchange effects into the mass and spin densities. In general, the formulation is much simpler than present formulations of many-body theory. The simplification comes from the fact that the model wavefunction contains only four independent variables, which can be mapped onto a basis set in real space. We have presented a proof-of-concept implementation and shown that it agrees well with the KS approach. 
 
 Given the orbital-free nature of this approach, it offers the possibility of truly linearly scaling DFT calculation; it will be possible to model systems of millions of atoms, bringing the biological domain into the range of full electronic structure calculations.

\begin{acknowledgments}
 The authors acknowledge EPSRC funding for the UKCP consortium, grant No. EP/K013610/1. This work was also supported by the North East Centre for Energy Materials (NECEM). Finally, this research made use of the Rocket High-Performance Computing service at Newcastle University.
\end{acknowledgments}

% \section*{References}
%  \bibliography{/home/tom/Desktop/refs}{}
% \bibliography{ref}{}
% \bibliographystyle{/home/tom/Desktop/latex/science}
% \bibliographystyle{/home/tom/Desktop/latex/rsc}
% \bibliographystyle{/home/tom/Desktop/latex/achemso}
\end{document}